\relax
\documentclass[letterpaper,man,apacite]{apa6} 
\usepackage[english]{babel}
\usepackage[utf8x]{inputenc}
\usepackage{amsmath}
\usepackage{graphicx}

\usepackage{color}
\newcommand{\rev}[1]{#1}

\title{Empirica: a virtual lab for high-throughput macro-level experiments}

\shorttitle{Empirica}

\author{
 Abdullah Almaatouq,\textsuperscript{\rm 1} 
 Joshua Becker,\textsuperscript{\rm 2} 
 James P. Houghton,\textsuperscript{\rm 1} 
 Nicolas Paton,\textsuperscript{\rm 1}\\
 Duncan J. Watts,\textsuperscript{\rm 3} and 
 Mark E. Whiting\textsuperscript{\rm 3}\vspace{10mm}}

\affiliation{
 \textsuperscript{\rm 1}Massachusetts Institute of Technology\\
 \textsuperscript{\rm 2}University College London\\
 \textsuperscript{\rm 3}University of Pennsylvania
}

\authornote{
Corresponding authors: \url{amaatouq@mit.edu} and \url{npaton@mit.edu}\\
Authors listed alphabetically by last name.
}

\abstract{
 Virtual labs allow researchers to design high-throughput and macro-level experiments that are not feasible in traditional in-person physical lab settings. Despite the increasing popularity of online research, researchers still face many technical and logistical barriers when designing and deploying virtual lab experiments. While several platforms exist to facilitate the development of virtual lab experiments, they typically present researchers with a stark trade-off between usability and functionality. We introduce Empirica: a modular virtual lab that offers a solution to the usability-functionality trade-off by employing a ``flexible defaults'' design strategy. This strategy enables us to maintain complete ``build anything'' flexibility while offering a development platform that is accessible to novice programmers. Empirica's architecture is designed to allow for parameterizable experimental designs, reusable protocols, and rapid development. These features will increase the accessibility of virtual lab experiments, remove barriers to innovation in experiment design, and enable rapid progress in the understanding of distributed human computation.
}

\begin{document}
\maketitle

\begin{figure}[tb]
 \centering
 \includegraphics[width=.7\columnwidth]{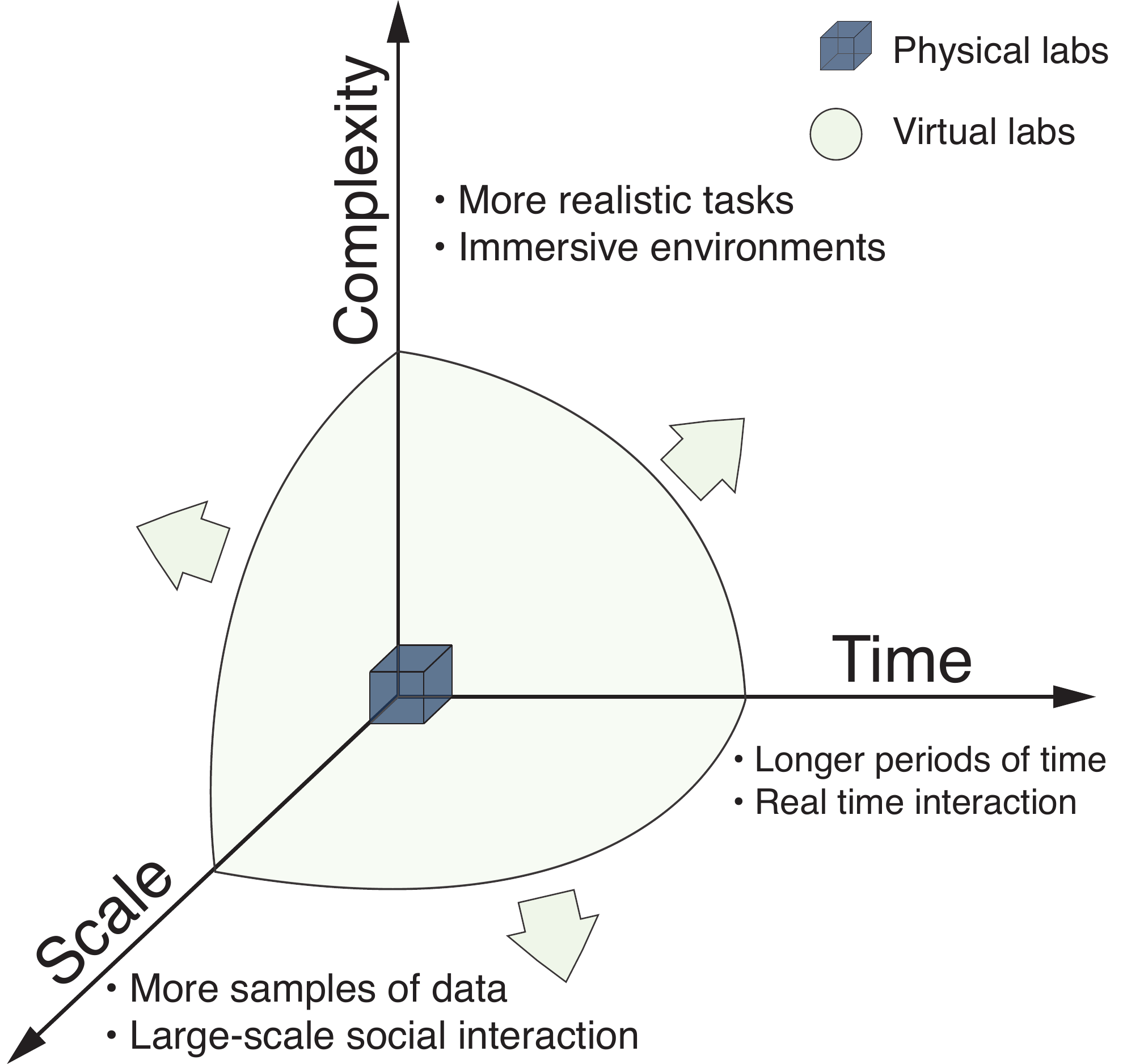}
 \caption{Schematic of the design space of lab experiments. Whereas many real-world social processes and phenomenon involve large numbers of people interacting in complex ways over long time intervals (days to years), physical lab experiments are generally constrained to studying individuals or small groups interacting in relatively simple ways over short time intervals (e.g., less than one hour). The potential of virtual lab experiments is that, in relaxing some of the constraints associated with in-person experiments, they can expand the accessible design space for social, behavioral, and economic experiments.}
 \label{fig:virtual_labs}
\end{figure}

Laboratory experiments are the gold standard for the study of human computation because they allow careful examination of the complex processes driving information processing, decision-making, and collaboration. Shortly after the World Wide Web had been invented, researchers began to employ ``virtual lab'' experiments, in which the traditional model of an experiment conducted in a physical lab is translated into an online environment~\cite{musch2000brief, Horton_Rand_Zeckhauser_2011,Mason_Suri_2012,reips2012using, Paolacci_Chandler_Ipeirotis_2010}. Virtual labs are appealing on the grounds that, in principle, they relax some important constraints on traditional lab experiments that arise from the necessity of physically co-locating human participants in the same room as the experimenter. Most obviously, virtual environments can accommodate much larger groups of participants than can fit in a single physical lab. However, as illustrated in Figure~\ref{fig:virtual_labs}, virtual lab experiments can also run for much longer intervals of time (e.g., days to months rather than hours) \rev{than is usually feasible in a physical lab} and can also exhibit more complex (e.g., complex network topologies, multifactor treatments) and more \rev{digitally} realistic designs. Finally virtual experiments \textit{can} be run faster and more cheaply than physical lab experiments, allowing researchers to explore more of the design space for experiments, with corresponding improvements in the replicability and robustness of findings.

Unfortunately, the potential of virtual lab experiments has thus far been limited by the often substantial up-front investment in programming and administrative effort required to launch them, effort that is often not transferable from one experiment to the next. An important step towards lowering the barrier to entry for researchers has therefore been the development of general-purpose virtual lab platforms (e.g., Qualtrics, jsPsych, nodeGame, oTree, lab.js). These platforms perform many of the functions of a virtual lab (e.g., data management, assignment to conditions, message handling) without the logic specific to a given experiment. In doing so, however, these platforms also present researchers with a trade-off between usability and flexibility. While some platforms provide graphical user interfaces (GUI) that are accessible to researchers with little or no programming experience, they achieve their usability by limiting the experiment designer to predetermined research paradigms or templates. In contrast, other platforms provide unlimited ``build anything'' functionality but require advanced programming skills to implement. As a result of this trade-off, many scientifically interesting virtual laboratory experiments that are theoretically possible remain prohibitively difficult to implement in practice.

A platform that maintains both usability and functionality will support methodological advancement in at least two high-priority areas. First, a highly usable platform is necessary for designing and administering \textit{high-throughput experiments} in which researchers can run, in effect, thousands of \rev{experimental conditions} that systematically cover the parameter space of a given experimental design. A legacy of the traditional lab model is that researchers typically identify one or a few theoretical factors of interest, and focus their experiment on the influence of those factors on some outcome behavior. Selectivity in conditions to be considered is sensible when only small numbers of participants are available. However, when many more participants are available, there is an opportunity to run many more conditions, and it is no longer necessary to focus on those that researchers believe a priori to be the most informative. In principle, researchers can define a set of dimensions along which the experiment can vary, and then a process can be used to generate and sample the set of conditions to be used in the experiment~\cite{balietti2018fast, mcclelland1997optimal}. For example, this approach was taken in the Choice Prediction Competitions, where human decision-making was studied by automatically generating over 100 pairs of gambles following a predefined algorithm~\cite{erev2017anomalies, plonsky2019predicting}. Recent work took advantage of the larger sample sizes that can be obtained through virtual labs to scale up this approach, collecting human decisions for over 10,000 pairs of gambles~\cite{bourgin2019cognitive}. The resulting data set can be used to evaluate models of decision-making and is at a scale where machine learning methods can be used to augment the insights of human researchers~\cite{agrawal2020scaling}. Also, there is still a lot of room to develop other kinds of experimental designs that are optimized for the high-throughput environment created by virtual labs. For example, one can navigate the increasingly large spaces of possible conditions and stimuli by making use of adaptive designs that intelligently determine the next conditions to run~\cite{balietti2020optimal, suchow2016rethinking, balandat2020botorch}. In order to make such experiments feasible, researchers need a platform that enables ``experiment-as-code,'' in which experiment design, experiment administration, and experiment implementation are separated and treated as code (where each can be formally recorded and replicated). This process allows for parameterizable designs, algorithmic administration, reusable protocols, reduced cost, and rapid development.

A second high priority in social science is the implementation of \textit{macro-level experiments} in which the unit of analysis is a collective entity such as a group~\cite{becker_network_2017,whiting2020parallel}, market~\cite{salganik2006experimental}, or an organization~\cite{valentine2017flash} comprising dozens or even hundreds of interacting individuals. As we move up the unit of analysis from individuals to groups, new questions emerge that are not answerable even with a definitive understanding of individual behavior~\cite{schelling2006micromotives}. At its most ambitious, macro-level experimentation offers a new opportunity to run experiments at the scale of societies. Previously, researchers who wanted to run experiments involving the interaction of hundreds of thousands of people only had the opportunity to do so in the context of field experiments. While this approach to experimentation is valuable for providing a naturalistic setting, it has major weaknesses in that such experiments are hard to replicate and typically provide only a single sample. Macro-level lab experiments typically require the design of complex tasks and user interfaces, the ability to facilitate synchronous real-time interaction between participants, and the coordination, recruitment, and engagement of a large number of participants for the duration of the experiment. Implementing large-scale macro experiments remains challenging in the absence of a virtual laboratory designed with multi-participant recruitment, assignment, and interaction as a core principle. Furthermore, running experiments that are both \textit{high-throughput} and \textit{macro-scale} requires a platform that simultaneously offers high usability while also maintaining a ``build anything'' functionality. 

To promote these methodological goals, Empirica offers a reusable, modular platform that facilitates rapid development through a ``flexible default'' design. This design provides a platform that is accessible to individuals with basic JavaScript skills but allows advanced users to easily override defaults for increased functionality. Empirica employs design features intended to aid and promote high-throughput and macro-scale experimentation methodologies. For example, the platform explicitly separates experiment design and administration from implementation, promoting the development of reliable, replicable, and extendable research by enabling ``experimentation-as-code.'' This modular structure encourages strategies such as multifactor~\cite{Almaatouq_Noriega-Campero_Alotaibi_Krafft_Moussaid_Pentland_2020}, adaptive~\cite{Letham_Karrer_Ottoni_Bakshy_2019,balietti2020optimal,Paolacci_Chandler_Ipeirotis_2010, balandat2020botorch}, and multiphase experimentation designs~\cite{mao2017resilient,almaatouq2020collective}, which dramatically expand the range of experimental conditions that can be studied. Additionally, the platform provides built-in data synchronization, concurrency control, and reactivity to natively support multi-participant experiments and support the investigation of macro-scale research questions. Empirica requires greater technical skill than GUI platforms, a design choice that responds to the emerging quorum of computational social scientists with moderate programming skills. Thus Empirica is designed to be ``usable'' for the majority of researchers while maintaining uncompromised functionality, i.e., the ability to build anything that can be displayed in a web browser.

After reviewing prior solutions, this paper provides a technical and design overview of Empirica. We then discuss several case studies in which Empirica was successfully employed to address ongoing research problems, and discuss the methodological advantages of Empirica. We conclude with a discussion of limitations and intended directions for future development. Throughout this paper, we will refer to ``games'' (experimental trials) as the manner in which ``players'' (human participants or artificial bots) interact and provide their data to researchers. This usage is inspired by the definition of human computation as ``games with a purpose''~\cite{von_Ahn_Dabbish_2008}, although many of the tasks would not be recognized as games as such.

\section{Related Work}
\subsection{Virtual Lab Participants}
It has long been recognized that the internet presents researchers with new opportunities to recruit remote participants for behavioral, social, and economic experiments~\cite{grootswagers2020primer}. For instance, remote participation allows researchers to solve some of the issues that limit laboratory research, such as (1) recruiting more diverse samples of participants than are available on college campuses or in local communities~\cite{reips2000web, berinsky2012evaluating}; (2) increasing statistical power by enabling access to larger samples~\cite{awad2018moral,reips2000web}; and (3) facilitating longitudinal and other multiphase studies by eliminating the need for participants to repeatedly travel to the laboratory~\cite{almaatouq2020collective,reips2000web}. The flexibility around time and space that is afforded by remote participation has enabled researchers to design experiments that would be difficult or even impossible to run in a physical lab.

Arguably the most common current strategy for recruiting online participants involves crowdsourcing services~\cite{Horton_Rand_Zeckhauser_2011,Mason_Suri_2012}. The main impact of these services has been to dramatically reduce the \rev{cost per participant in lab studies}, resulting in an extraordinary number of publications in the past decade. Unfortunately, a limitation of the most popular platforms such as Amazon Mechanical Turk \rev{or TurkPrime~\cite{litman2017turkprime}} is that they were designed for simple labeling tasks that can typically be completed independently and with little effort by individual ``workers'' who vary widely in quality and persistence on the service~\cite{goodman2013data}. Moreover, Amazon's terms of use prevent researchers from knowing whether their participants have participated in similar experiments in the past, raising concerns that many Amazon ``turkers'' are becoming ``professional'' experiment participants~\cite{chandler2014nonnaivete}. In response to concerns such as these, services such as Prolific\footnote{\url{www.prolific.co}}~\cite{palan2018prolific} have adapted the crowd work model to accommodate the special needs of behavioral research. For example, Prolific offers researchers more control over participant sampling and quality as well as recruiting participants who are intrinsically motivated to contribute to scientific studies. 

In addition to crowdsourcing services, online experiments have attracted even larger and more diverse populations of participants who participate voluntarily out of intrinsic interest to assist in scientific research. For example, one experiment collected almost forty million moral decisions from over a million unique participants in over 200 countries~\cite{awad2018moral}. Unfortunately, while the appeal of ``massive samples for free'' is obvious, all such experiments necessarily rely on some combination of gamification, personalized feedback, and other strategies to make participation intrinsically rewarding~\cite{Hartshorne_de_Leeuw_Goodman_Jennings_O_Donnell_2019}. As a consequence, the model has proven hard to generalize to arbitrary research questions of interest. 

\subsection{Existing Virtual Lab Solutions}
While early online experiments often required extensive up-front customized software development, a number of virtual lab software packages and frameworks have now been developed that reduce the overhead associated with building and running experiments. As a result, it is now easier to implement designs in which dozens of individuals interact synchronously in groups~\cite{arechar2018conducting,almaatouq2020collective,whiting2019did} or via networks~\cite{becker_network_2017}, potentially comprising a mixture of human and algorithmic agents~\cite{ishowo2019behavioural,traeger2020vulnerable,shirado2017locally}.

Virtual lab solutions can be roughly grouped by their emphasis on usability or functionality. Here we describe free or open-source tools that allow synchronous, real-time interaction between participants, leaving aside tools such as jsPsych~\cite{de_Leeuw_2015}, lab.js~\cite{henninger2019lab}, and Pushkin~\cite{Hartshorne_de_Leeuw_Goodman_Jennings_O_Donnell_2019} that do not explicitly support multi-participant interactions as well as commercial platforms such as Testable, Inquisit, Labvanced~\cite{finger2017labvanced}, and Gorilla~\cite{anwyl2020gorilla}.

Platforms such as WEXTOR~\cite{Reips_Neuhaus_2002}, Breadboard~\cite{mcknight2016breadboard}, and LIONESS~\cite{giamattei2019lioness} provide excellent options for individuals with little-to-no coding experience. These platforms allow researchers to design their experiments either directly with a graphical user interface (GUI) or via a simple, proprietary scripting language. However, while these structures enable researchers to quickly develop experiments within predetermined paradigms, they constrain the range of possible interface designs. These platforms do not allow the researcher to design ``anything that can run in a web browser.''

On the other hand, many excellent tools including oTree~\cite{Chen_Schonger_Wickens_2016}, nodeGame~\cite{Balietti_2017}, Dallinger,\footnote{\url{docs.dallinger.io}} and TurkServer~\cite{Mao_Chen_Gajos_Parkes_Procaccia_Zhang_2012} offer high flexibility in experiment design. However, this flexibility comes at the expense of decreased usability, as these tools require significant time and skill to employ. They are flexible precisely because they are very general, which means additional labor is required to achieve any complete design. 

\section{Empirica}
The Empirica platform\footnote{\url{empirica.ly}} is a free, open-source, general-purpose virtual lab platform for developing and conducting synchronous and interactive human-participant experiments. The platform implements an Application Programming Interface (API) that allows an experiment designer to devote their effort to implementing participant-facing views and experiment-specific logic. In the background, Empirica handles the necessary but generic tasks of coordinating browser-server interactions, batching participants, launching games, and storing and retrieving data.

Experiments are deployed from a GUI web interface that allows the researcher to watch the experiment progress in real time. With no installation required on the participant’s part, experiments can run on any web browser including desktop computers, laptops, smartphones, and tablets~(See Appendix~\ref{appx:admin_interface}).

Empirica is designed using a ``flexible default'' strategy: the platform provides a default structure and settings that enable novice JavaScript users to design an experiment by modifying pre-populated templates; at the same time, unlimited customization is possible for advanced users. The goal of this design is to develop a platform that is accessible to researchers with modest programming experience --- the target user is the typical computational social science researcher --- while maintaining a ``build anything'' level of flexibility. 

Empirica has an active and growing community of contributors, including professional developers, method-focused researchers, question-driven social scientists, and outcome-oriented professionals. Although Empirica is under steady development, it has already been used to build (at least) 31 experiments by more than 18 different research teams across 12 different institutions, generating at least 12 manuscripts between 2019 and 2020~\cite{feng2019exploring, pescetelli2019collective, Becker_Porter_Centola_2019, becker2019crowd, Almaatouq_Noriega-Campero_Alotaibi_Krafft_Moussaid_Pentland_2020, houghton_2020, becker2020network, almaatouq2020collective, noriega2020screening, feng2020towards, houghton2020interdependent, guilbeault2020probabilistic, jahani2020exposure}.

\subsection{System Design}

\begin{figure}[tb]
 \centering
 \includegraphics[width=.7\columnwidth]{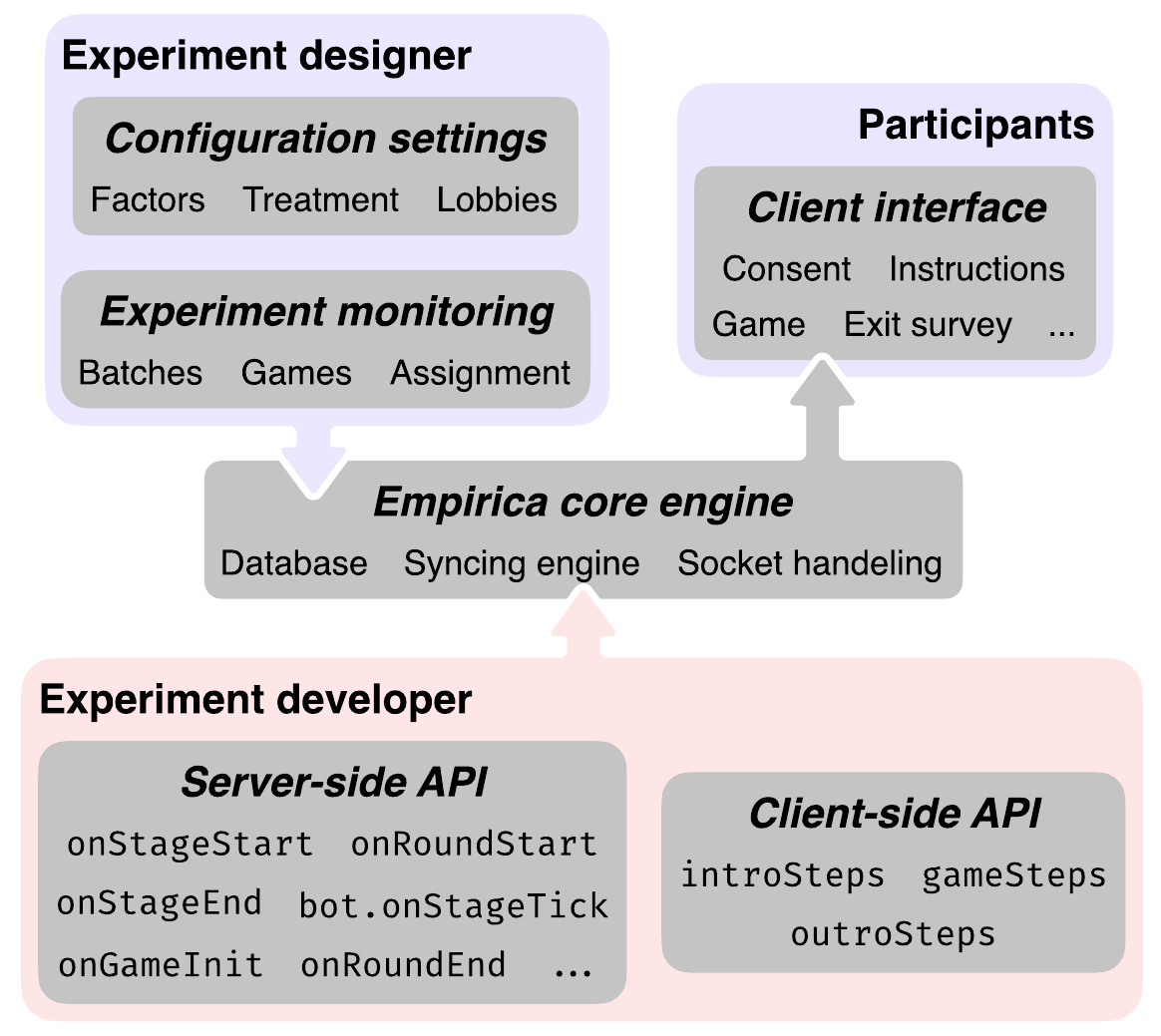}
 \caption{Empirica provides a scaffolding for researchers to design and administer experiments via three components: (1) Server-side callbacks use JavaScript to define the running of a game through the client-side and server-side API; (2) the client-side interface uses JavaScript to define the player experience; and (3) the GUI admin interface enables configuration and monitoring of experiments (see Appendix~\ref{appx:admin_interface}). These components are all run and connected by the Empirica core engine.}
 \label{fig:empirica_overview}
\end{figure}

Empirica's architecture was designed from the start to enable real-time multi-participant interactions, although single-player experiments are easy to create as well. The API is purposefully concise, using a combination of data synchronization primitives and callbacks (i.e., event hooks) triggered in different parts of the experiment. The core functionality is abstracted by the platform: data synchronization, concurrency control, reactivity, network communication, experiment sequencing, persistent storage, timer management, and other low-level functions are provided automatically by Empirica. As a result, researchers can focus on designing the logic of their participants' experience (see Figure~\ref{fig:empirica_overview} for an overview).

To initiate development, Empirica provides an experiment scaffold generator that initializes an empty (but fully functioning) experiment and a simple project organization that encourages modular thinking. To design an experiment, researchers separately configure the client (front end), which defines everything that participants experience in their web browser, thus defining the experimental treatment or stimulus, and the server (back end), which consists of callbacks defining the logic of an experimental trial. The front end consists of a sequence of five modules: consent, intro (e.g., instructions, quiz), lobby, game, and outro (e.g., survey). The lobby\footnote{Because participants usually do not arrive at precisely the same time, and also because different participants require more or less time to read the instructions and pass the quiz, Empirica implements a virtual ``lobby'' feature. While waiting in the lobby, participants receive information about how much time they have been waiting and how many other players are still needed for the experiment to start.} serves the purpose of starting a new experimental trial when specific criteria are met (e.g., a certain number of participants are simultaneously connected) and it is automatically generated and managed by Empirica according to parameters set in the GUI. The researcher need only modify the intro, outro, and game design via JavaScript. The back end consists of callbacks defining game initialization, start and end behavior for rounds and stages, and event handlers for changes in data states.

Empirica structures the game (experimental trial) as players (humans or artificial participants) interacting in an environment defined by one or more rounds (to allow for ``repeated'' play); each round consists of one or more stages (discrete time steps), and each stage allows players to interact continuously in real time. Empirica provides a timer function which can automatically advance the game from stage to stage, or researchers can define logic that advances games based on participant behavior or other conditions.

As Empirica requires some level of programming experience for experiment development, the platform accommodates the possibility that different individuals may be responsible for designing, programming, and administering experiments. To support this division of labor, Empirica provides a high-level interface for the selection of experimental conditions and the administration of live trials. From this interface, experiment administrators can assign players to trials, manage participants, and monitor the status of games. Experiment designers can configure games to have different factors and treatments, and export or import machine-readable \textit{YAML}\footnote{YAML Ain't Markup Language (YAML) is a data serialization language designed to be human-friendly and work well with modern programming languages~\cite{ben2009yaml}.} files that fully specify entire experiment protocols (i.e., the data generation process) and support replication via experiment-as-code. Experiment configuration files can also be generated programmatically by researchers wishing to employ procedural generation and adaptive experimentation methods to effectively and efficiently explore the parameter space.

The ultimate test of an experiment's design is that it is able to evaluate its target theory. In addition to creating artificial players to use as part of an experiment, Empirica's ``bot'' API also allows users to perform full integration tests of their experiment. By simulating the complete experiment under all treatments with simulated participants, the experiment designer can ensure that their as-implemented design matches their expectation.

\begin{figure*}[tb]
 \centering
 \includegraphics[width=\columnwidth]{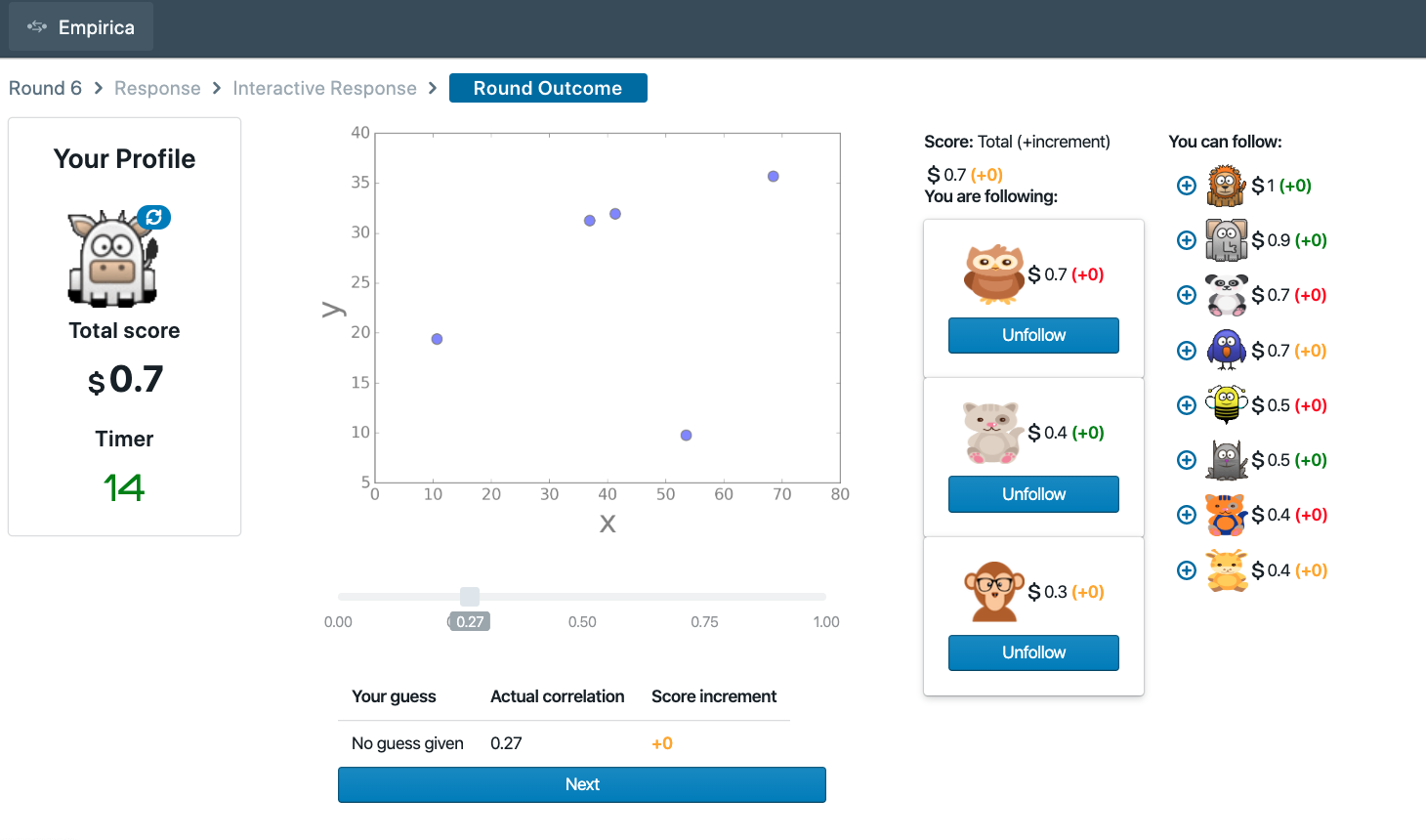} 
 \caption{This screenshot of the ``Guess the Correlation Game'' shows the view that participants use to update their social network in the dynamic network condition with full feedback (i.e., as opposed to no feedback or only self-feedback). In all of the experimental condition, the maximum number of outgoing connections was set to 3 and the group size is set to 12.} The interface uses reactive and performant front-end components.
 \label{fig:guess_the_correlation_game}
\end{figure*}

\begin{figure*}[tb]
 \centering
 \includegraphics[width=\columnwidth]{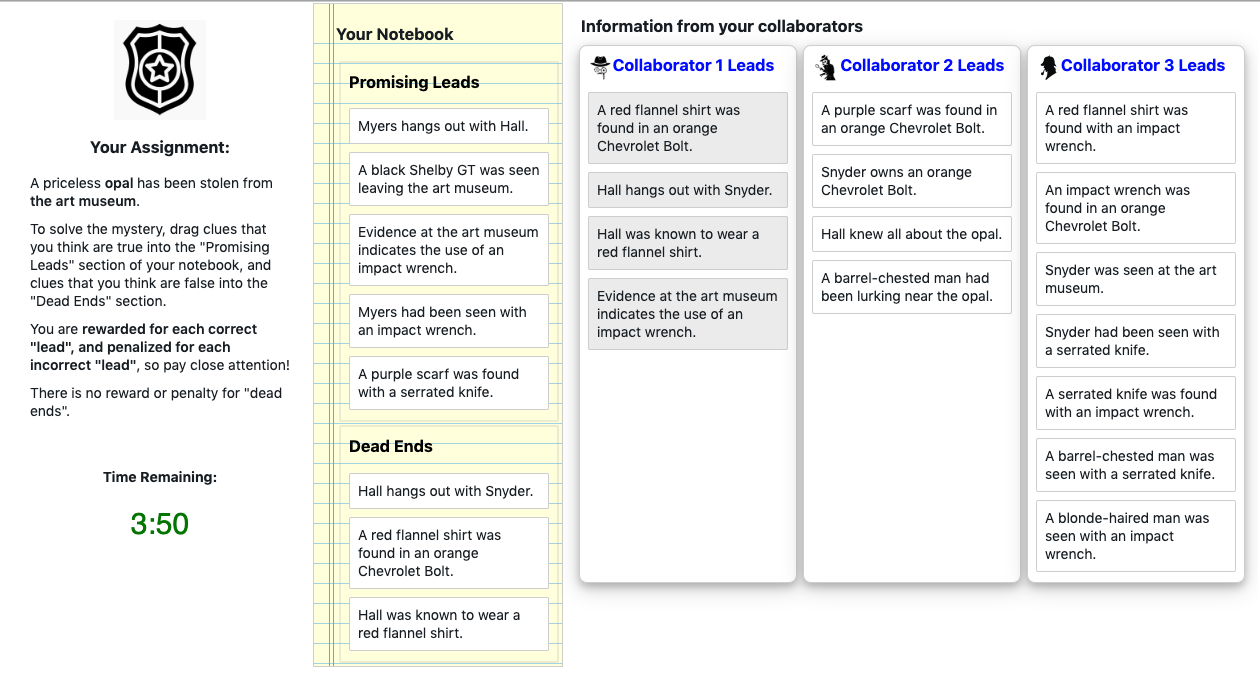}
 \caption{This screenshot of the ``Detective Game'' shows the view that participants use to categorize mystery clues as either Promising Leads (which are shared with their social network neighbors) or Dead Ends (which are not). The interface uses reactive and performant front-end components.}
 \label{fig:detective_game}
\end{figure*}

\begin{figure*}[tb]
 \centering
 \includegraphics[width=\columnwidth]{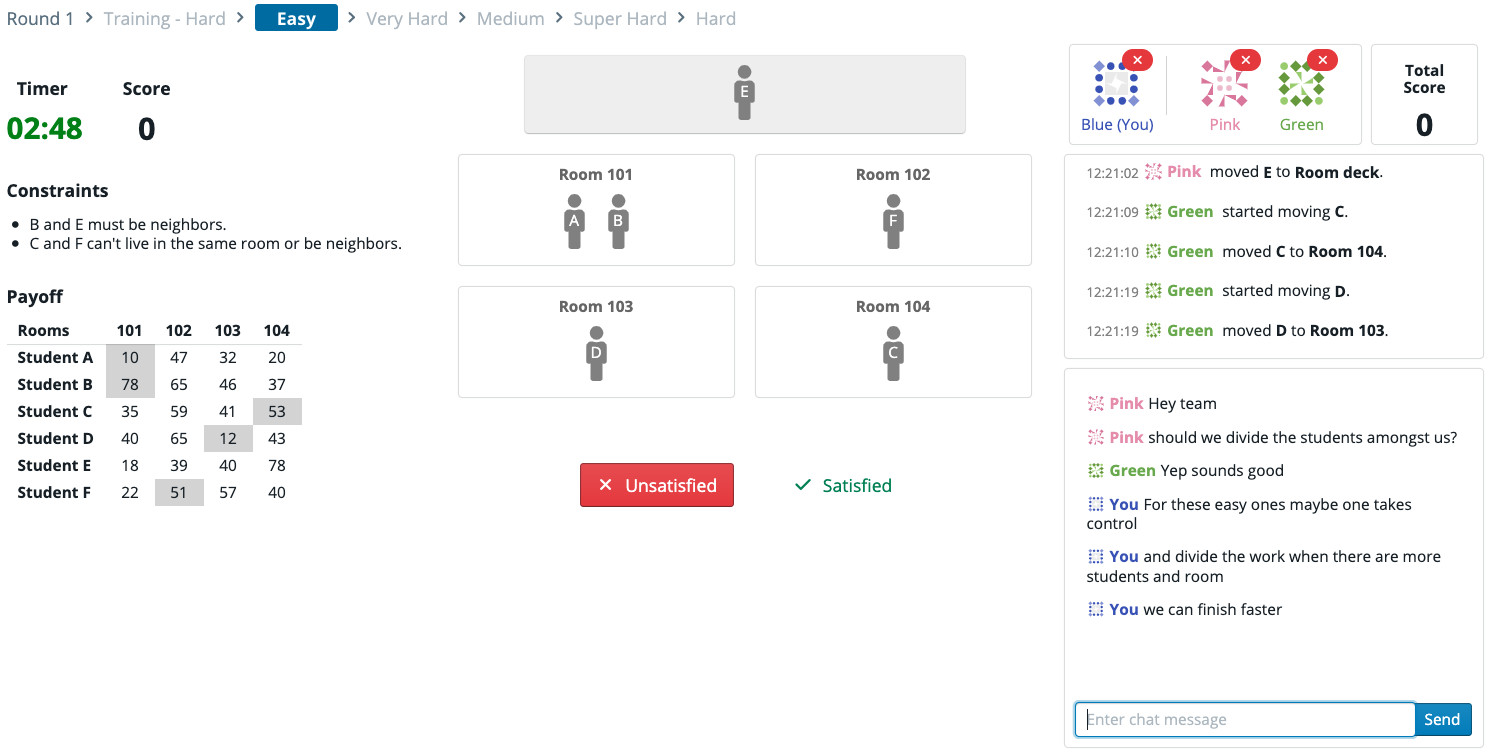}
 \caption{This screenshot shows the ``Room Assignment'' task. The real-time interaction, the ability to assign students to rooms in parallel, and text-based chat employs default features and interaction components provided by Empirica.}
 \label{fig:room_assignment}
\end{figure*}

\begin{figure*}[tb]
 \centering
 \includegraphics[width=\columnwidth]{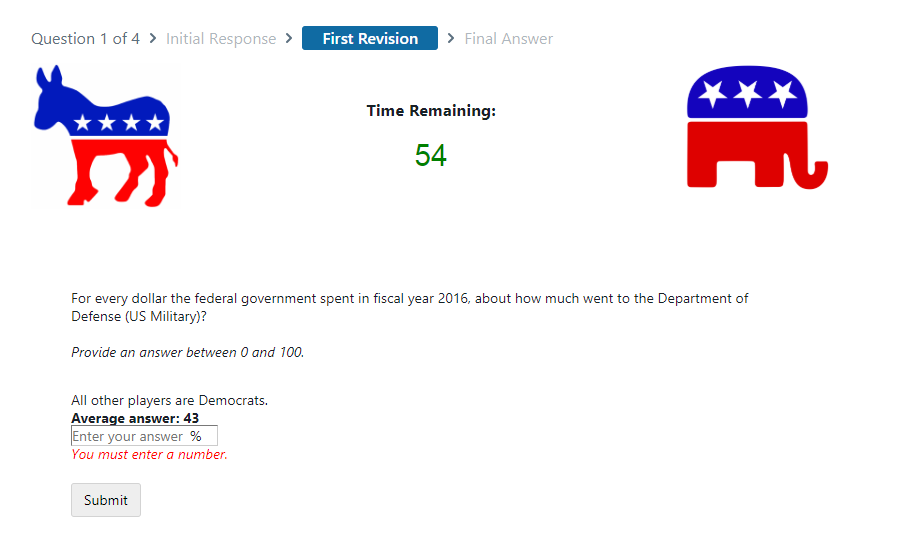}
 \caption{This screenshot shows the second stage of the first round of the revised ``Politics Challenge'' estimation task. The illustrated breadcrumb feature employs customized default UI elements provided from Empirica, and the timer was employed without modification.}
 \label{fig:politics_challenge}
\end{figure*}

\subsection{Implementation}
Empirica is built using common web development tools. It is based on the \textit{Meteor}\footnote{\url{www.meteor.com}} application development framework and employs JavaScript on both the front end (browser) and the back end (server). \textit{Meteor} implements tooling for data reactivity around the MongoDB database, WebSockets, and RPC (remote procedure calls). \textit{Meteor} also has strong authentication, which secures the integrated admin interface (see Appendix~\ref{appx:admin_interface}). Experiment designers will not need to be familiar with \textit{Meteor} to use the Empirica platform. Only those who wish to contribute to the development of Empirica and contribute to the codebase will need to use \textit{Meteor}.

The front end is built with the UI framework \textit{React},\footnote{\url{reactjs.org}} which supports the system's reactive data model. Automatic data reactivity implemented by Empirica alleviates the need for the experiment designer to be concerned with data synchronization between players. \textit{React} has a vibrant and growing ecosystem, with many resources from libraries to online courses to a large talent pool of experienced developers, and is used widely in production in a variety of combinations with different frameworks~\cite{fedosejev2015react, wieruch2017road}. For Empirica, \textit{React} is also desirable because it encourages a modular, reusable design philosophy. Empirica extends these front-end libraries by providing experimenter-oriented UI components such as breadcrumbs showing experiment progression, player profile displays, and user input components (e.g., Sliders, text-based Chat, Random Dot Kinematogram). These defaults reduce the burden on experiment designers while maintaining complete customizability.It is important to note that it is up to the experiment developer to follow the best practices of UI development that are appropriate for their experiment. \rev{For instance, behavioral researchers interested in timing-dependent procedures should be cautious when developing their UIs and should test the accuracy and precision of the experimental interface~\cite{garaizar2019best}.} Similarly, browser compatibility will depend on which \textit{React} packages are being used in the particular experiment. 

Empirica's back end is implemented in \textit{node.js}\footnote{\url{nodejs.org}} --- a framework for developing high-performance, concurrent programs~\cite{tilkov2010node}. Callbacks are the foundation of the server-side API. Callbacks are hooks where the experiment developer can add custom behavior. These callbacks are triggered by events of an experiment run (e.g., \texttt{onRoundStart}, \texttt{onRoundEnd}, \texttt{onGameEnd}, \texttt{...}). The developer is given access to the data related to each event involving players and games and can thus define logic in JavaScript that will inspect and modify this data as experiments are running.

This design allows Empirica to reduce the technical burden on experiment designers by providing a data interface that is tailored to the needs of behavioral lab experiments. The developer has no need to interact with the database directly. Rather, Empirica provides simple accessors (\texttt{get}, \texttt{set}, \texttt{append}, \texttt{log}) that facilitate data monitoring and updating. These accessor methods are available on both the front end and the back end. All data is scoped to an experiment-relevant construct such as game, player, round, or stage. Data can also be scoped to the intersection of two constructs, e.g., a player and a game object: player.round and player.stage which contain the data for a player at a given round or stage. The accessor methods are reactive, meaning that data is automatically saved and propagated to all players. Empirica's front end and back end are connected over WebSocket (a computer communications protocol), where a heartbeat (or ping) continuously monitors the connection and allows the server to determine if the client is still responsive. On the player side, on disconnection, the client will passively attempt to reconnect with a session identifier stored in the browser's local storage. From the experiment developers' point of view, they can configure the experiment to: (1) continue without the missing player; (2) cancel the entire experimental trial; (3) pause the experimental trial (currently being implemented for future release); or (4) implement a custom behavior (e.g., a combination of 1-3).

Another ease-of-use feature is that an Empirica experiment is initialized with a one-line command in the terminal (Windows, macOS, Linux) to populate an empty project scaffold. A simple file structure separates front-end (client) code from back-end (server) code to simplify the development process. Because Empirica is built using the widely adopted \textit{Meteor} framework, a completed experiment can also be deployed with a single command to either an in-house server or to a software-as-a-service platform such as \textit{Meteor Galaxy}. Additionally, Empirica provides its own simple open-source tool to facilitate deploying Empirica experiments to the cloud for production.\footnote{\url{github.com/empiricaly/meteor-deploy}} This facilitates iterative development cycles in which researchers can rapidly revise and redeploy experiment designs.

Empirica is designed to operate with online labor markets such as \rev{Prolific} or other participant recruitment sources (e.g., volunteers, in-person participants, classrooms).

\section{Case Studies}
\rev{Throughout its development, Empirica has been used in the design of cutting-edge experimental research. Below, we illustrate Empirica’s power and flexibility in four examples, each of which highlights a different functionality.
}

\subsection{Exploring the parameter space: Dynamic social networks and collective intelligence}
\rev{The ``Guess the Correlation''~\cite{Almaatouq_Noriega-Campero_Alotaibi_Krafft_Moussaid_Pentland_2020}\footnote{The source code for the ``Guess the Correlation'' experiment can be found at
\url{github.com/amaatouq/guess-the-correlation}} game was developed to study how individual decisions shape social network structure ultimately determining group accuracy (Figure~\ref{fig:guess_the_correlation_game}). 
}

\rev{In this game, participants were tasked with estimating statistical correlations from a visual plot of two variables (such as height and weight). For each image, participants first guessed individually and could then update their guesses while seeing other participants guesses and updates in real-time. Between rounds, participants could see feedback on each other's accuracy and could add/drop people from the social network that determined whose answers were shown.
}

\rev{In this game, participants were tasked with estimating statistical correlations from a visual plot of two variables (such as height and weight). For each image, participants first guessed individually and could then update their guesses while seeing other participants guesses and updates in real-time. Between rounds, participants could see feedback on each other's accuracy and could add/drop people from the social network that determined whose answers were shown.
}

\rev{The final publication reported seven experimental conditions with three varied levels of social interaction and four levels of performance feedback, and found that a variety of subtle changes could dramatically influence macro-scale group outcomes. The results show that even subtle changes in the environment can lead to dramatically different macro-scale group outcomes despite any micro-scale changes in individual experience.
}

\subsection{Real-time interaction at scale: A large-scale game of high-speed ``Clue''}
\rev{The ``Detective Game''~\cite{houghton_2020}\footnote{The source code for the ``Detective Game'' experiment can be found at
\url{github.com/JamesPHoughton/detective\_game\_demo}} examined the effect of belief interaction on social contagion. In the game, teams of 20 players worked together to solve a mystery by exchanging clues. To coordinate recruitment and ensure proper randomization, the experimenter planned to recruit up to 320 participants to participate in each block of games. 
}

\rev{However, this number of simultaneous participants is two orders of magnitude larger than in typical behavioral experiments, and the participants needed to interact in real time. The interface showed players when peers updated their beliefs and when they added clues around to their ``detective's notebook,'' as shown in Figure~\ref{fig:detective_game}. The experimenter needed a platform with short load times, high-performance display libraries, and imperceptible latency at scale. At the same time, their code needed to be readable enough for academic transparency. 
}

\rev{The experimenter used Empirica’s ``flexible default'' design and modular API to quickly evaluate a number of open-source display libraries, selecting from the multiplicity of modern web tools those which best supported the experiment. They then used Empirica's ``bot'' API to simulate player's actions in the game, testing that the back-end could provide the low-latency coordination between client and server crucial to the game's performance. The experiment confirmed theoretical predictions that belief interaction could lead to social polarization.
}

\subsection{Two-phase experiment design: Distributed human computation problems}
\rev{The ``Room Assignment'' game~\cite{almaatouq2020collective}\footnote{The source code for the ``Room Assignment'' experiment can be found at \url{github.com/amaatouq/room-assignment}} explored how factors such as task complexity and group composition allow a collaborating team to outperform its individual members. The task consisted of a ``constraint satisfaction and optimization'' problem in which $N$ ``students'' were to be assigned to $M$ ``dorm rooms'', subject to constraints and preferences (Figure~\ref{fig:room_assignment}). 
}

\rev{Unlike many group experiments, this study required the same group of participants to perform the task twice. In the first round, participants needed to perform the task individually so that their individual skill level, social perceptiveness, and cognitive style could be measured. Then, in the second round, participants would be assigned to collaborate in teams using Empirica's included chatroom plugin chat, a standard empirica plugin\footnote{The Chat component is available at \url{github.com/empiricaly/chat}}. This simple design enabled researchers to measure task performance for independent and interacting groups while controlling communication, group composition, and task complexity.
}

\rev{The experimenters used Empirica’s careful participant data management and flexible randomization architecture to reliably match the same subject pool across the two phases of this experiment and to coordinate the large block-randomized design. While this may have been possible with other platforms, Empirica's admin interface made these considerations as simple as making selections from a drop-down list.
}

\subsection{Rapid-turnaround replication: Echo chambers and belief accuracy}
\rev{The ``Estimation Challenge'' experiment~\cite{Becker_Porter_Centola_2019}\footnote{The source code for the ``Estimation Challenge'' experiment can be found at
\url{github.com/joshua-a-becker/politics-challenge}} tested how politically-biased echo chambers shape belief accuracy and polarization. Participants answered factual questions (such as ``How has the number of unauthorized immigrants living in the US changed in the past 10 years?'') before and after observing answers given by other participants. The experimenters found that collective intelligence can increase accuracy and decreased polarization despite popular arguments to the contrary.
}

\rev{This experiment was implemented using a custom platform in partnership with a third-party developer and generated an ad-hoc social network to determine how information flowed among participants. After submitting these results for publication, the reviewers expressed concern that the experiment design did not fully capture the effects of a politicized environment. The experimenters were given 60 days to revise and resubmit their paper.
}

\rev{Revising the original interface in the time available or rehiring the original developer would have required skills or monetary resources not available to the project. Using Empirica, they were able to replicate the initial experiment with a modified user interface to address the questions posed by reviewers, as seen in Figure~\ref{fig:politics_challenge}. The new interface was designed, constructed, and tested in approximately two weeks. This experiment required negligible alteration from the prepopulated Empirica scaffolding beyond customizing the visual design and introductory steps, demonstrating the capability of flexible defaults.
}

\section{Discussion}
\subsection{Ethical considerations}
As with any human subjects research, virtual lab experiments are subject to ethical considerations. These include (but are not limited to) pay rates for participants~\cite{whiting2019fair}, data privacy protection~\cite{birnbaum2004human}, and the potential psychological impact of stimulus design. While most of these decisions will be made by the researchers implementing an experiment using Empirica, we have adopted a proactive strategy that employs default settings designed to encourage ethical experiment design. As one example, the initial scaffolding generated by Empirica includes a template for providing informed consent, considered a bare minimum for ethical research practice. The scaffolding also includes a sample exit survey which models inclusive language; e.g., the field for gender is included as a free-text option. To encourage privacy protection, Empirica by default omits external identifiers when exporting data to prevent leaking of personal information such as email addresses or Amazon Turk account identifiers. 

\subsection{Limitations and future developments}
As with other leading computational tools, Empirica is not a static entity, but a continually developing project. This paper reflects the first version of the Empirica platform, which lays the groundwork for an ecosystem of tools to be built over time. Due to its design, modules that are part of the current platform can be switched out and improved independently without rearchitecting the system. Indeed it is precisely because Empirica (or for that matter, any experiment platform) cannot be expected to offer optimal functionality indefinitely that this modular design was chosen.

The usability-functionality trade-off faced by existing experiment platforms is endemic to tightly integrated ``end-to-end'' solutions developed for a particular class of problems. By moving toward an ecosystem approach, Empirica has a chance to resolve this trade-off. As such, future development of Empirica will include the development of a set of open standards that defines what this encapsulation (service/component) is, how to communicate with it, and how to find and use it. 

The use of the ``ecosystem'' as a design principle presents several opportunities for operational efficiency.
\begin{itemize}
\item An ecosystem will allow the reuse of software assets, in turn lowering development costs, decreasing development time, reducing risk, and leveraging existing platform investments and strengths. 
\item The individual components of the ecosystem will be loosely coupled to reduce vendor/provider lock-in and create a flexible infrastructure. As a result, the individual components of the ecosystem will be modular in the sense that each can be modified or replaced without needing to modify or replace any other component because the interface to the component remains the same. The resulting functional components will be available for end users (i.e., researchers) to amalgamate (or mashup) into situational, creative, and novel experiments in ways that the original developers may not originally envision. 
\item The functional scope of these components will allow for the possibility to directly define experiment requirements as a collection of these functional components, rather than translating experiment requirements into lower-level software development requirements. As a result, the ecosystem will abstract away many of the logistical concerns of running experiments, analogous to how cloud computing has abstracted away from the management of technical resources for many companies.
\end{itemize}

By distancing ourselves from a monolithic approach, and adopting a truly modular architecture with careful design of the low-level abstractions of experiments, we hope Empirica will decouple flexibility from ease-of-use and open the door to an economy of software built around conducting new kinds of virtual labs experiments.

\section{Conclusion}
Empirica provides a complete virtual lab for designing and running online lab experiments taking the form of anything that can be viewed in a web browser. The primary philosophy guiding the development of Empirica is the use of ``flexible defaults,'' which is core to our goal of providing a ``do anything'' platform that remains accessible to a typical computational social scientist. In its present form, Empirica enables rapid development of virtual lab experiments, and the researcher need only provide a recruitment mechanism to send participants to the page at the appropriate time. Future versions of Empirica will abstract the core functionality into an ecosystem that allows the development and integration of multiple tools including automated recruitment. This future version will also maintain as a ``tool'' the current Empirica API, continuing to enable the rapid development of experiments.

\section{Open Practices Statements}
Empirica is entirely open-source and in active development. The codebase is currently hosted on Github.\footnote{\url{github.com/empiricaly}} Documentation and tutorial videos are available at \url{docs.empirica.ly}. We encourage readers who are interested in the software to contribute ideas or code that can make it more useful to the community. 

\section{Acknowledgments}
The authors are grateful to all the persons who have contributed to the development of Empirica over the years. A special thanks to the super contributor Hubertus Putu Widya Primanta Nugraha. We also thank the users of Empirica for suggestions for improvement and reporting bugs. We were supported by a strong team of advisors including Iyad Rahwan, Matthew Salganik, Alex ‘Sandy’ Pentland, Alejandro Campero, Niccolò Pescetelli, and Joost P Bonsen. The authors gratefully acknowledge the Alfred P. Sloan Foundation (G-2020-13924) and the James and Jane Manzi Analytics Fund for financial support.

\bibliography{bib}

\appendix
\section{Empirica Admin Interface}\
\label{appx:admin_interface}
\rev{View of the admin interface provided by Empirica. Panel (A) shows the experiment ``monitoring'' view. Panel (B) shows the experiment ``configuration'' view.}
\begin{figure}
 \centering
 \includegraphics[width=0.9\columnwidth]{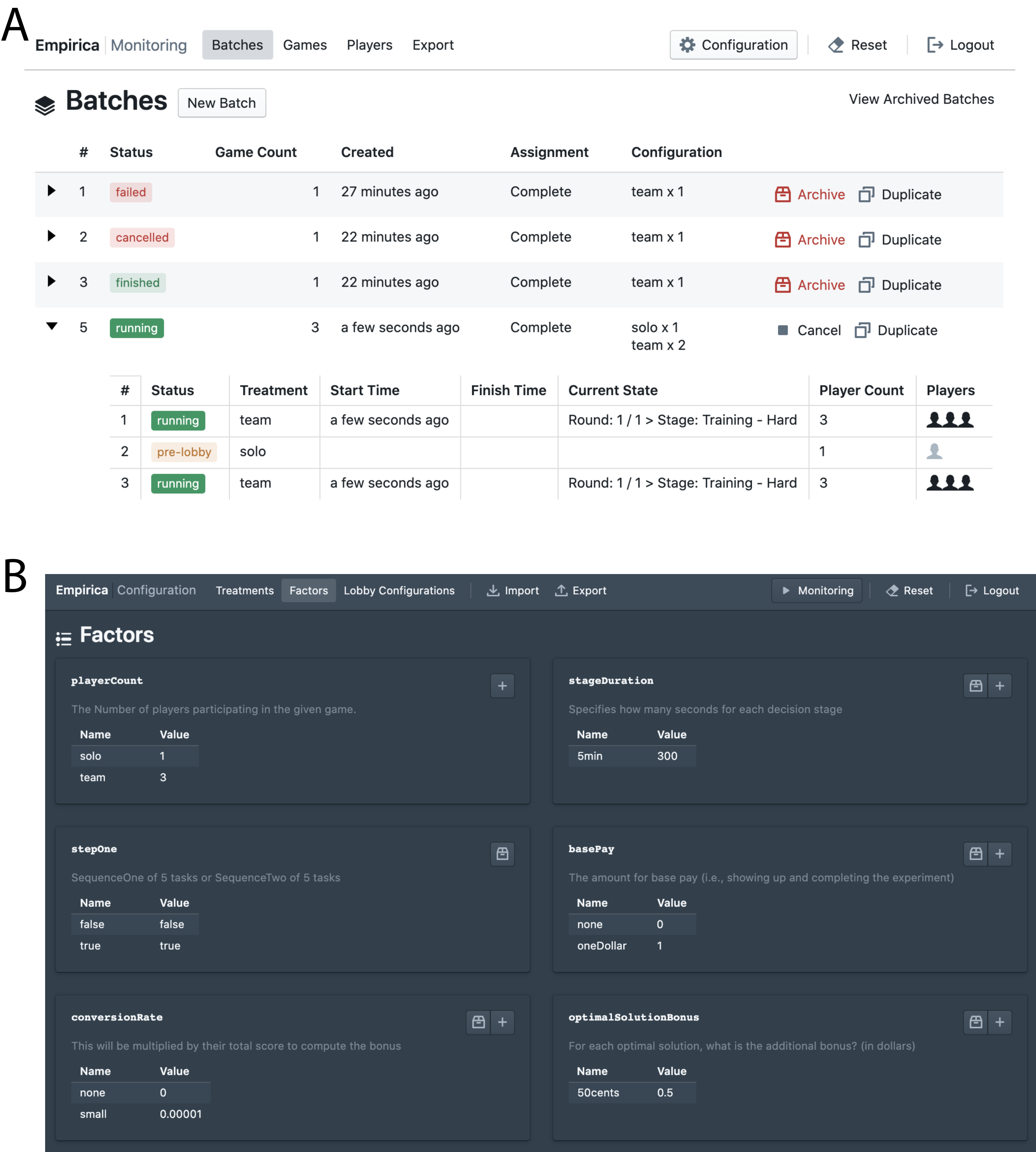}
\end{figure}

\end{document}